\pdfoutput=1

\documentclass[aps,showpacs,preprintnumbers,amsmath,amssymb,
eqsecnum, twocolumn, tightenlines
]{revtex4}

\usepackage{graphicx}

\newcommand{\be}{\begin{eqnarray}}
\newcommand{\ee}{\end{eqnarray}}

 \newcommand{\gsim}{\mathrel{\hbox{\rlap{\lower.55ex \hbox {$\sim$}}
                   \kern-.3em \raise.4ex \hbox{$>$}}}}
\newcommand{\lsim}{\mathrel{\hbox{\rlap{\lower.55ex \hbox {$\sim$}}
                   \kern-.3em \raise.4ex \hbox{$<$}}}}

\begin{document}


\title{  Jet/Fireball Edge should be observable! }
\author { Edward Shuryak}
\address { Department of Physics and Astronomy, State University of New York,
Stony Brook, NY 11794}
\date{\today}

\begin{abstract}
Shock/sound propagation from the quenched jets have well-defined front, separating the fireball into regions which are and are not
affected. While even for the most robust jet quenching observed this increases  local temperature and flow of ambient matter  by only few percent at most,   strong radial flow increases the contrast between the two regions so that the difference should be well seen
in particle spectra at some $p_t$,
perhaps even on event-by-event basis. We further show that the effect  comes mostly from certain ellipse-shaped 1-d curve, the intercept of three 3-d surfaces,
 the Mach cone history, the timelike and spacelike freezeout surfaces. We further suggest  that this ``edge" is already seen in  an event   released by ATLAS collaboration.
\end{abstract}
\maketitle

\section{Introduction of the idea}

Observation of jets at RHIC are limited to the transverse energy in the range 20-30 GeV, which is quite
 difficult because of large and strongly fluctuating background.  Therefore
most of the studies has been based on the two and three-hadron correlation functions. 
Furthermore, for  hadrons  mostly studies their transverse momenta  are in the range of several GeV, where contributions from  hard jets 
and the tail
of hydrodynamical flow  is hard to separate uniquely. 
With the ``Little Bang" arriving at LHC in November 2010, the situation has changed
since at LHC much higher energy jets are available, for which triggering on jets works well.
The first glimpse of what is to come has been spectacularly demonstrated
by ATLAS collaboration in their first heavy ion paper \cite{ATLAS_jets} devoted to jet quenching.
Now the trigger jets have the transverse energy $E_\perp > 100 \, GeV$: excellent calorimeter of ATLAS
make standard jet finding algorithms to work well. The distribution over lost energy were found to be very sensitive to centrality, and for central collisions
 significant part of jet energy is lost,  in some events  completely.

In the present paper we turn to discussion of 
perturbations of the ``Little Bang" by the energy deposited by jets.
As evidenced by the enhanced radial and elliptic flows \cite{ALICE_v2} , overall hydrodynamical picture seem to work at LHC as well as at RHIC. 
  Once the
energy is deposited into the medium by the jet, it will result in
shock/sound perturbations in the shape of  the Mach cone   \cite{Casalderrey-Solana:2004qm,Horst}, similar e.g. to  lightning and thunder.
The present paper points out that very strong radial flow allows one to significantly simplify the problem, by focussing only the overlap of the
Mach (lifeline) 3d surface with the time-like and the space-like freezeout surfaces.


\begin{figure}[t]
\includegraphics[width=7cm]{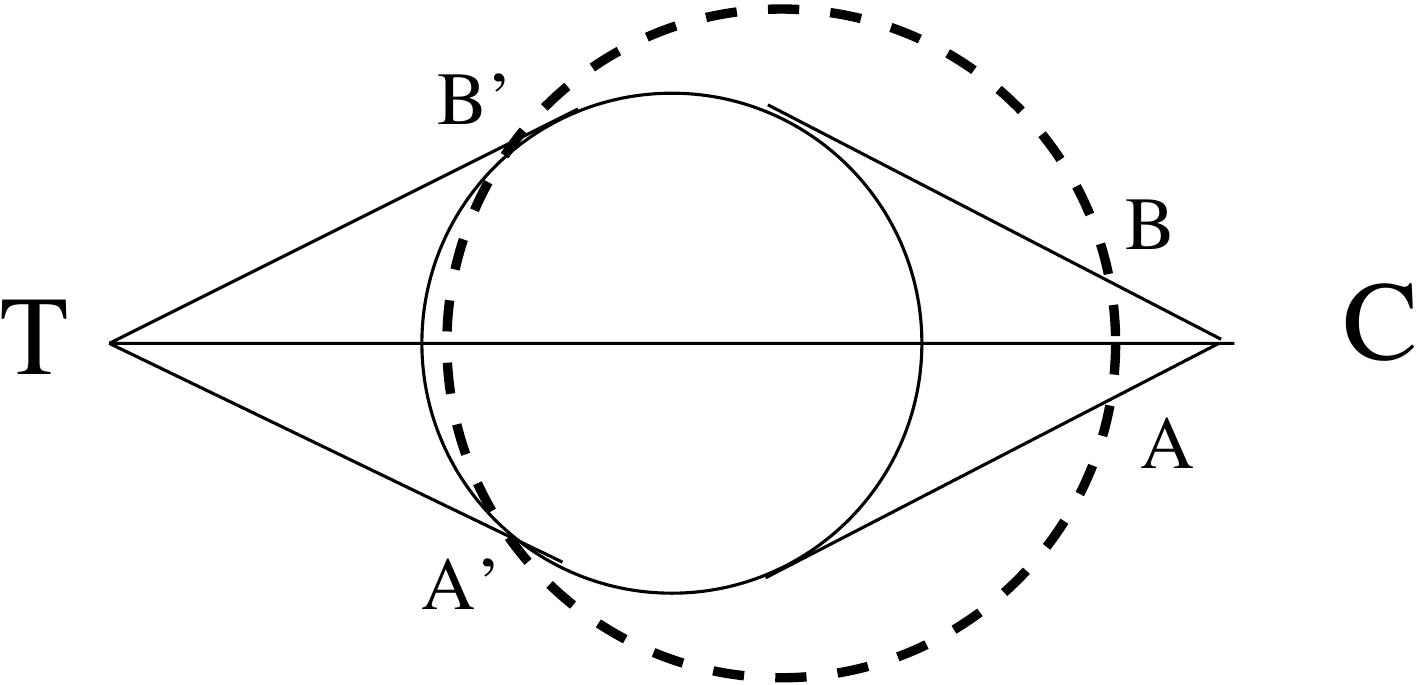}
\includegraphics[width=8cm]{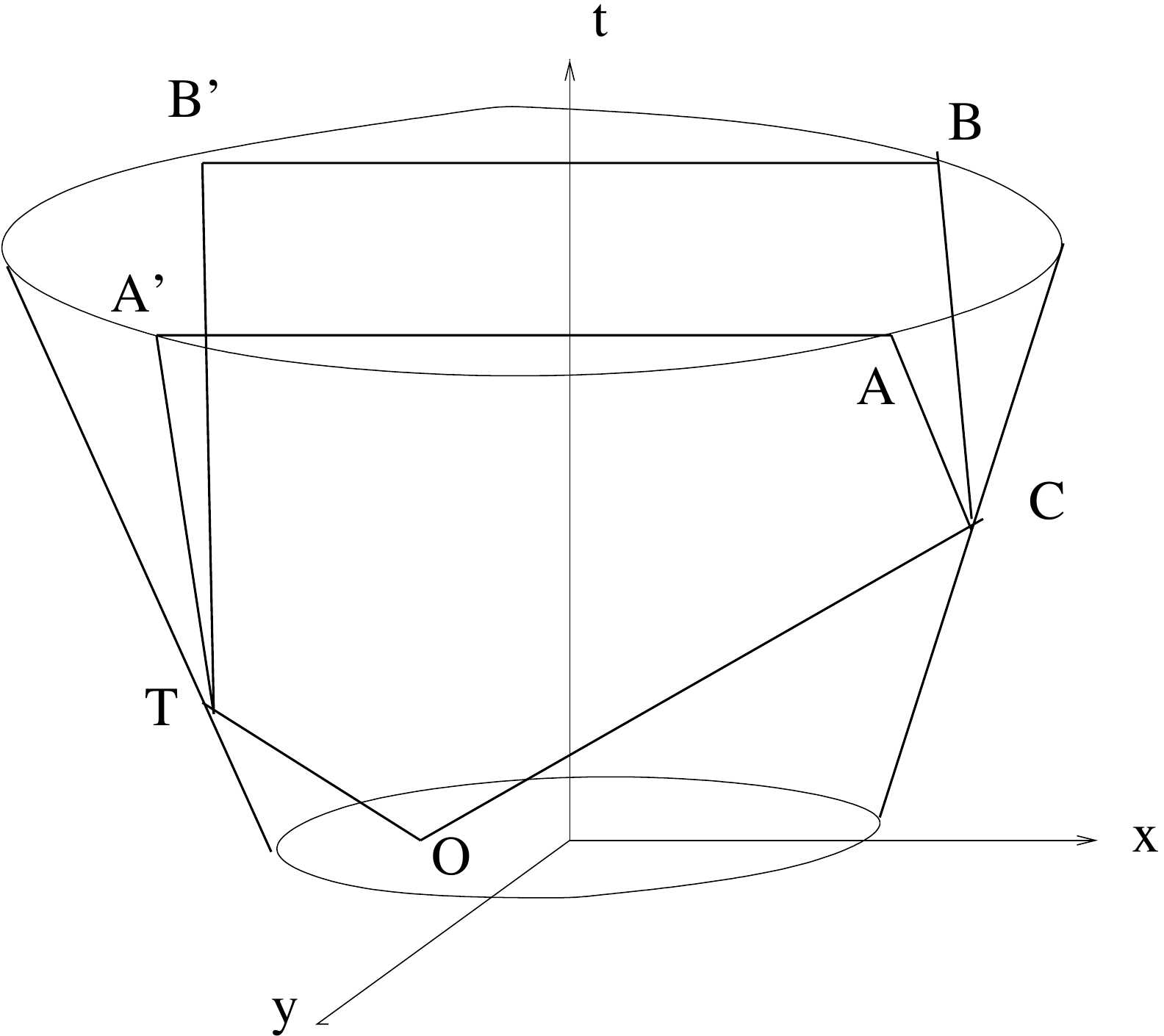}
  \caption{Schematic shape of the Mach surface in the transverse $x,y$ plane at $z=0$ and fixed time (upper plot), as well as its shape
  in 3d including the (proper longitudinal) time (lower plot). Mach surface $\sigma_M$ is made of two parts, $OCAA'T$ and $OCBB'T$. For more explanations see text.   }
  \label{fig_cone}
\end{figure}

The main idea to be presented is based on  two very simple geometrical observations: \\
(i)  whatever 
complicated distorsions of the  Mach cone in exploding matter may appear, 
 the  observed spectra come mostly from 
its intersection with the fireball space-time boundary known as a
 freezeout surface. \\
 (ii) Furthermore, because of the Hubble-like nature of the radial flow, the effect is strongly peaked 
 at the intersection of all three surfaces, the Mach surface $\sigma_M$,  the timelike and spacelike freezeout surfaces, denoted by  $\sigma_t ,\sigma_s$ respectively.
 
 Since each 3-d surface is one equation in 4-d space-time, the intersection of three of them are (two) 4-3=1-d lines, $\epsilon_C,\epsilon_T$, to be specified below. It is those lines which we call the 
jet/fireball edges: its size will also be estimated and compared with the data. Hydrodynamical causality would require that only a patch of matter inside it
is affected by the jet, while that outside it is unperturbed. We will show below that the contrast between those two regions can be experimentally
observable.

The geometry of the problem is schematically explained in Fig.\ref{fig_cone}. Its upper part is a snapshot at some time and some longitudinal coordinate, taken to be zero $z=0$, 
of the hydrodynamical perturbation
in the transverse plane. In infinite homogeneous matter   two back-to-back jets depositing energy into the medium  create
two cones tangent to a sphere, shown by the continuous lines. The shock/sound speed is assumed to be  roughly constant, about half of the speed of light.
If matter is present only onside the fireball, approximated by the (dashed) sphere, only the part
inside it actually exists. The intersection of this sphere with the affected matter happens at 4 points, indicated as $A,B$ for a companion jet $C$ and $A',B'$ for a 
trigger jet $T$. The so called ``trigger bias" leads to a trigger jet passing less matter than the companion jet: thus the picture is  left-right asymmetric. 
(We assume central collisions and a jet with  transverse momentum  in x direction,  travelling along the diameter, thus  up-down symmetry
of the plot.)

The lower plot of   Fig.\ref{fig_cone} shows this picture in 3d, still at the same $z=0$ but now including the so called longitudinal proper time $t=\sqrt{t_{lab}^2-z^2}$,
where $t_{lab}$ is the laboratory time. This time runs upward, and the upper ellipse schematically represent the time-like part of the freezeout surface, $\sigma_t$,
approximated by the constant time surface, $t=t_f$. The lower ellipse is the ``initiation time surface", and the conical surface connecting them is our approximation
to the space-like part of the freezeout surface, $\sigma_t$. Inside the region hydrodynamics is assumed to be valid,
as usual, whole outside secondaries freestream to the detector. The points $A,B,A',B'$ have the same meaning as in the upper plot, which can be seen as 
corresponding to its $\sigma_t$ face. Since we only show the 3d picture ($z=0$) three surfaces we speak about are 2-dimensional, and their overlap 
is 3-3=0 dimensional, reduced to 4 points $A,B,A',B'$.

For obvious reason we do not show 4-d plot, but this is not needed. Adding the longitudinal $z$ direction is simple, it sill promote the edge into two ellipses,
one having points $A,B$ on it and one having $A',B'$. Those will be called the $edges$ $\epsilon_C,\epsilon_T$ of the companion and trigger jets, respectively. Since
trigger-bias force the companion jet to deposit much larger amount of energy, the former one has much larger chance to become visible.

\section{Further details} 
 The general expression for a spectrum is thermal spectrum boosted by the flow $u^\mu$ and  integrated over the 3d freezeout surface (the Cooper-Fry formula)
\be  {dN\over d^3p}= \int_{\sigma_t+\sigma_s} d\Sigma_\mu p^\mu exp\left[ - {p_\nu u^\nu \over T}\right] \ee
Let us first quantify the second part of the idea (ii),  that  the intercept with $both$ surfaces $\sigma_t,\sigma_s$  is the most visible one.
(we will only show it for $\sigma_t$ part, similar argument holds for the $\sigma_s$ part as well.)

 Focusing on the exponent and using (for simplicity) nonrelativistic approximation for the flow
($u_0=cosh y_\perp \approx 1, u_r=sinh y_\perp \approx  y_\perp$) and Hubble parameterization
for the radial flow near the surface of the fireball as $ y_\perp(r,\tau)\approx H (R-\delta r)$ one can simplify the relevant exponential factor into
\be exp\left( - {p_t\over T}  RH {\delta r \over R} \right) \ee
The first ratio, $p_t/T$ should be taken as large as possible, remaining at the same time
in the validity region of the hydrodynamical description of the spectra. Let us say take $p_t$ 
to be 2 GeV for RHIC and 3 GeV for LHC. The second factor $RH$, the maximal value of the
transverse rapidity of the flow, is about 0.7 and 0.8, respectively.  Let
the freezeout temperature be $T_f=0.12 \, GeV$. One finds $exp(-12\delta r / R)$ for RHIC
and even  $exp(-20\delta r / R)$ for LHC, which means that only a small vicinity of the rim
$\delta r / R=0.1-0.05$ is ``experimentally visible".  

The Mach surface $\sigma_M$ surrounds the matter which is affected by the jet. Let us provide a simple (upper limit) estimate of how different
this matter is from the unperturbed ambient matter. Using mid-rapidity ALICE multiplicity $dN_{ch}/d\eta\approx 1584$ of the charged particle, we
multiply it by 3/2 to include neutrals and get $dN/d\eta\sim 2400$. Since the rapidity width of the region affected by a jet has $\Delta \eta \sim 1$, this multiplicity
can be directly compared with the ``extra particles" originated from the jet. At deposited $E_\perp\sim 100 \,GeV$ this number is about $N_{extra}\sim 200$,
provided they are equilibrated with flowing medium completely, an increase of about 8\%. Since multiplicity scales as $T^3$, the increase of the temperature
(if homogeneous) is about $\delta T/T\sim 2.7 \%$ only, which does not look like much. And yet, this effect is so much amplified by the radial
flow that it should be easily observable, in a specially tuned region of the spectra. 

At the freezeout the matter density is roughly constant (e.g. twice larger multiplicity at LHC relative to RHIC leads to twice large HBT volume, as shown by ALICE \cite{ALICE_HBT}). So,    $N_{extra}$ particles need about   8\% of extra volume. Assuming that longitudinal expansion is still rapidity-independent, it means increasing
the transverse area, or increasing the freezeout radius by the square root of it, or 4\% in our example. The Hubble law of expansion then tell us that it will increase flow
velocity linearly with $r$, or also by  4\%. The boost exponent however can easily increase the contrast  to be  as large as 100\%, for example
\be exp[({p_t u_t \over T_f}) {\delta u_t \over u_t} ]\sim exp(20*0.04)\sim 2.2  \ee
(using the same parameters  as in the example above).   
  

So far our discussion is completely geometrical in nature: all we have used above is that
jets move with a speed of light and shock/sound with a speed $c_s$ (although depending on the density/temperature, but the same for all events in the same centrality bin).  Note that neither jet energy
nor its fraction deposited are  important. The difference between RHIC and LHC
collisions only come from a somewhat different multiplicity and timing. 

So far, we have ignored many complications. 
The speed of the shock/sound is not in fact constant and
depend on the amplitude a bit and also 
 has a dip near the QCD
phase transition:  this should somewhat deform the Mach surface. 
Another issue (see below) is the interaction between the wave and the flow.

So far we have ignored the dynamics of jet quenching itself,
on which the magnitude of the observed signal  depends. Let us thus briefly mention the evolution of current views on the
 quenching mechanism. 
The very first measurements at RHIC have provided the magnitude of the 
attenuation of the hadron spectra, known as the $R_{AA}(p_t)$. Although
 the magnitude of suppression is quite large, up to factor 5, perturbative
 models were able to reproduced it. However next RHIC discoveries put pQCD explanations into doubt.  Single leptons, originating
 from $c,b$ quark decays, 
show equally small $R_{AA}^{c,b}$, which is hard to explain perturbatively. Another issue,
  pointed out in \cite{Shuryak:2001me,Drees:2003zh}, is large 
 angular asymmetry of 
the jet quenching  incompatible with  models for which $-dE/dx$ is proportional to the matter density. The 
value of 
$v_{2}(p_{t}>6\, GeV)= <cos(2\phi)>$ is wrong by a factor 2.
 (Note high $p_t$: it should not  be confused with the
elliptic flow!) 
One possible solution \cite{Liao:2008dk} to this puzzle is a nontrivial dependence of quenching rate on matter density, with
 enhancement in the near-$T_c$ region.
Another solution came from   the strong coupling
(AdS/CFT) framework, which predicts  the energy loss  \cite{Chesler}  
\be %
-{dE \over dx} \sim T^{4}x^{2}\,, \label{eqn_ADS_DEDX}
\ee %
with an extra peak at the very end near the stopping point. 
While this result may look similar to the perturbative BDMCS result \cite{Baier:1994bd} 
 $-{dE \over dx} \sim
T^{3}x$, the timing predicted by those two regimes are in fact completely different. 
 Strong coupling scenario (\ref{eqn_ADS_DEDX})  effectively
shifts most of the  jet quenching to its latest moments, and therefore \cite{Marquet:2009},
leads to $v_2(p_t>6\, GeV)$ large enough to reproduce the PHENIX and STAR data.
This last scenario, if true, provides an independent dynamical reason why only the jet edge (defined above) rather than the whole cone, is dominant in the observed spectra. 
(Note that in the discussion above we have only discussed observability
of the edge of the companion jet: scenario  (\ref{eqn_ADS_DEDX}) provides another
reason why the energy deposition of the trigger jet should be significantly smaller,
and thus its edge less visible.)

\section{The wave carried by the flow}
We defer complete treatment of the sound propagation on top of expanding fireball
to larger paper \cite{ShuryakStaig}, in which we will use a number of approximate
methods as well as complete separation of variables for the so called Gubser flow.   
In this paper we focus on the jet/fireball edge, and thus can 
 describe sound propagation in the ``geometric acoustics" approximation, see textbooks such as
 \cite{LL_fluid}. It uses the analogy between the Hamilton-Jacobi equation for
the particle and the wave sound equation, putting the Hamilton eqns of motion for
``phonons"  in the following general
form
\be %
{d\vec{r} \over dt}= {\partial \omega(\vec{k},\vec{r}) \over \partial \vec{k}}  \label{eqn_speed}\,,  \\
{d\vec{k} \over dt}=-{\partial \omega(\vec{k},\vec{r}) \over
\partial \vec{r}}\,, \label{eqn_force}
\ee %
driven by the (position dependent) dispersion relation
$\omega(\vec{k},\vec{r})$. For clarity, let us start with the
simplest non-relativistic case, namely with the small velocity of
the flow, $u \ll 1$. In this case the dispersion relation is
obtained from that in the fluid at rest by a local Galilean
transformation, so that for flow $\vec{u}(\vec r)$
\be %
\omega(\vec{k},\vec{r})=c_{s} k + \vec{k}\vec{u}(\vec r)\,.
\label{eqn_nonrel}
\ee %
%
I

Note that for a wave propagating normal to the flow
(in the azimuthal angle direction) this equation
 contains only the first term. (We neglect small 
elliptic and other higher angular harmonics or just consider central collisions). Thus
 the distance travelled by the sound is just the ``sound horizon"
$  H_s=c_s(t_f-t_C) $, where $t_C$ is proper time at which the jet lives the fireball, see Fig.1.  
In longitudinal direction one expects that the sound travels  longer distance, as it partly rides on
the longitudinal flow. The equation for the longitudinal position of the sound pulse looks  now
\be 
\dot{z}(t)=c_s+{z(t) \over \sqrt{t^2-z^2(t)}} \ee
It can either be solved numerically, or (with about 2\% accuracy) approximated nonrelativistically
(dropping $z$ in the denominator of the second term), which  leads to analytic
 solution
$ Z_s= c_s t_f ln({t_f \over t_C}) $
which by the time of final freezeout is greater than $H_s$ by about factor $Z_s/H_s\approx 1.4$. The shape of the jet edge is thus approximately an ellipsoid with the ratio
of radii $Z_s/H_s$. 


What happens if the jet stops inside the fireball? The upper plot in Fig.1 gets modified, the cone gets ``rounded" by a sphere centered at the stopping point. 
Its intersection with the two other freezeout surfaces can still be defined, as above.

Crude estimate of the azimuthal angle of the edge is given by $\Delta \phi= \pm {H_s\over R_f}$, the distance shock/sound travel after the jet left the fireball to the final fireball radius. 
It is of the order of one radian numerically: but specific values depend both on the length of the jet path inside the matter and on the wave speed.
The shock speed for QGP-to-QGP case we worked out for
 different  compression factors:  there is no place here to present this calculation,
 and we just 
say that in the  range of compression factors of interest shock rapidity remains close to 1/2.



\begin{figure}[t]
\includegraphics[width=8cm]{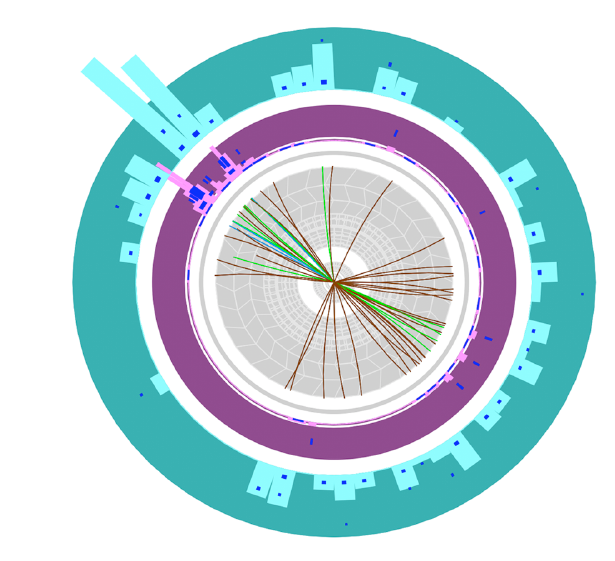}
  \caption{Azimuthal distribution of the transverse energy in one event from \cite{ATLAS_jets} ,  the inner part shows tracks with $p_\perp > 2.6 \, GeV$ and 
outer histogram the hadronic 
calorimeter cells, with thresholds $E>1\, GeV$.
 }
  \label{fig_ATLAS_event}
\end{figure}

\section{Experimental signatures}
As argued above, each jet which deposit certain amount of energy into hydrodynamical process effectively 
heats up matter insider the Mach surface, which results in extra radial flow in certain angular sector. We have argued 
that one should look at $p_t=2-3\, GeV$, a window where hydrodynamical effects are still dominant and the ``contrast" is the largest.
We expect to find sharp jump between the inside of this sector and the outside. Furthermore, there should be extra
peaks near the edge itself, corresponding to the ``frozen sound pulse". In principle, this should happen both for companion and trigger jets, although with larger 
magnitude in the former case.

This statement is of course statistical, and should be studied for a sample of events with close energy deposition.
 However, the predicted contrast is large enough to be perhaps seen in
single events.  One event display  shown in ATLAS paper is reproduced in  our Fig. 2; note that tracks $p_t$ and calorimeter  energy cuts are in the range we propose.
Indeed, tracks and calorimeter energy distributions
are very peculiar. First of all, they are  wide and not jet-like,
as has been observed in pp collisions. Second, the distributions are not Gaussian-like but flat, with clear sharp edges
beyond which there is no signal. We suggest that this is the 
 manifestation of the  ``edge" phenomenon discussed.
Last but not least, there are symmetrically placed small peaks near both edges: those presumably are due to  
 ``frozen sound waves". If one accounts for longitudinal (first order velocity) $\delta u$ in exponent of (2.1), their azimuthal angle is slightly larger than that of the geometric edge
 discussed above.
Although this display shows only one  event, being projected onto the azimuthal angle, in fact
the edge and near-edge enhancement should make an ellipse in the $\phi,\eta$ plan, as we discussed above. 

In summary, we suggest that events in which large energy is deposited by jets should develop sharp ``edges" in angular distribution of particles, best seen for  $p_t=2-3\, GeV$,
which are of geometric nature and thus nearly 
independent on the jet energy. We expect them to become a new experimental tool, as they should be noticeable  even on event-by-event basis. 

 {\bf Acknowledgments.} 
 This work was
supported in parts by the US-DOE grant DE-FG-88ER40388.

\end{document}